# BARD: A seamless two-stage dose optimization design integrating backfill and adaptive randomization

Yixuan Zhao[1], Rachael Liu[2], Jianchang Lin[2], Ying Yuan[3, *]

[1]Department of Biostatistics, University of Texas Health Science Center at Houston, TX,USA
[2]Department of Biostatistics, Takeda Pharmaceuticals, Boston, USA
[3]Department of Biostatistics, University of Texas MD Anderson Cancer Center, TX, USA
*Correspondence to: yyuan@mdanderson.org

Word Count: 5371

# Abstract

One common approach for dose optimization is a two-stage design, which initially conducts dose escalation to identify the maximum tolerated dose (MTD), followed by a randomization stage where patients are assigned to two or more doses to further assess and compare their risk-benefit profiles to identify the optimal dose. A limitation of this approach is its requirement for a relatively large sample size. To address this challenge, we propose a seamless two-stage design, BARD (Backfill and Adaptive Randomization for Dose Optimization), which incorporates two key features to reduce sample size and shorten trial duration. The first feature is the integration of backfilling into the stage 1 dose escalation, enhancing patient enrollment and data generation without prolonging the trial. The second feature involves seamlessly combining patients treated in stage 1 with those in stage 2, enabled by covariate-adaptive randomization, to inform the optimal dose and thereby reduce the sample size. Our simulation study demonstrates that BARD reduces the sample size, improves the accuracy of identifying the optimal dose, and maintains covariate balance in randomization, allowing for unbiased comparisons between doses. BARD designs offer an efficient solution to meet the dose optimization requirements set by Project Optimus, with software freely available at [www.trialdesign.org](www.trialdesign.org).

# Introduction

Conventionally, phase I dose-finding trials aim to identify the maximum tolerated dose (MTD) under the assumption that both toxicity and efficacy increase with dose. However, this paradigm poses challenges for targeted therapies and immunotherapies, where the monotonicity assumption often doesn't hold.[1, 2] For instance, when a targeted agent's binding is saturated before reaching the MTD, increasing the dose may not improve efficacy further. In such cases, a dose below the MTD may offer a better benefit-risk tradeoff by providing similar efficacy with lower toxicity and better tolerability.[3] Recognizing this issue, the United States Food and Drug Administration (FDA) launched Project Optimus[4] to reform the dose selection paradigm. This initiative shifts the focus of dose finding and selection from the MTD to the optimal biological dose (OBD), which delivers the optimal risk-benefit profile.

Numerous methods have been proposed to identify the OBD. Yuan et al (2016)[5] reviewed phase I-II trial designs and discussed critical topics in OBD identification. To better understand these various designs, Yuan et al. (2024)[6] classified them into two strategies: the efficacy-integrated strategy and the two-stage strategy. The efficacy-integrated strategy considers the risk-benefit tradeoff from the beginning of the trial and uses it to guide the dose finding. Examples of efficacy-integrated designs include the model-based design such as EffTox design and late-onset EffTox,[7, 8] and model-assisted designs such as BOIN12, BOIN-ET, and uTPI.[9, 10, 11] Efficacy-integrated designs are efficient for identifying the OBD and are most suitable for cases where the efficacy endpoint can be ascertained relatively quick and also patient population is expected to be similar between dose finding and subsequent phase II trials. A number of clinical trials applied this strategy to find the OBD using EffTox or BOIN12.[12, 13, 14]

The two-stage strategy refers to the approach that first performs dose escalation to identify the MTD and safe dose range, followed by a randomization stage where patients are randomized between two or more doses to further assess and compare the risk-benefit profiles of these doses for identifying the OBD. Compared to the efficacy-integrated strategy, the two-stage strategy is more flexible, allowing different populations in the two stages (e.g., all-comers in stage 1 and particular indications in stage 2). In addition, randomization also decreases heterogeneity and enables more unbiased comparisons between doses.[15] This approach has been described in the FDA's guidance on dose optimization.[16] Examples of trial designs using the two-stage strategy include the method by Hoering et al (2011),[17] DROID,[18] and U-BOIN.[19]

One limitation of two-stage designs is their requirement for a relatively large sample size due to their structured approach. In stage 1, typical sample sizes for dose escalation often range from 20 to 30 patients, depending on the number of doses (e.g., 4-5 doses). For stage 2, Yang et al. (2023)[20] recommended sample sizes of 20 to 40 patients per dose arm to achieve reasonable accuracy in identifying the OBD. As a result, the total sample size required is substantially larger than that in conventional dose-finding trials, leading to increased costs and longer development times.

To address this challenge, we propose a seamless two-stage design, BARD (Backfill and Adaptive Randomization for Dose Optimization), incorporating two key features to reduce sample size and shorten trial duration. The first feature of BARD is the integration of backfilling into stage 1 dose escalation, allowing additional patients to be treated (backfilled) at doses deemed safe and showing promising activity. This concurrent approach enhances patient enrollment and data generation without prolonging the trial duration, thereby better informing the determination of the MTD and OBD. The second feature seamlessly combines patients treated in stage 1 with those in stage 2, significantly reducing the overall sample size requirement. Integrating stage 1 patients, who are not randomized, into stage 2 may compromise covariate balance across doses. To mitigate this, we employ covariate-adaptive randomization in stage 2 to actively address potential imbalances in prognostic factors among stage 1 non-randomized patients. Our simulation study demonstrates that this approach achieves covariate balance comparable to that of fully randomized trials comprising only randomized patients.

The remainder of the paper is organized as follows: We propose the methodology and the BARD design in Section 2, describe a simulation study and sensitivity analysis in Section 3, and provide concluding remarks in Section 4.

# Method

BARD consists of two seamless stages: stage 1 conducts dose escalation with backfill, and stage 2 performs covariate-adaptive randomization. The objective of stage 1 is to establish the MTD and provide toxicity and efficacy data, as well as pharmacokinetics and pharmacodynamics data, to select the doses for stage 2 randomization. Depending on the dose escalation method used in stage 1, different versions of BARD can be constructed. We focus on the BOIN[21] and Bayesian logistic regression model (BLRM)[22] methods to illustrate the use of model-assisted and model-based dose escalation designs, respectively, while noting that our methodology can readily accommodate other dose escalation methods, such as the keyboard design[23] and continuous reassessment method (CRM).[24]

# Stage 1 Dose Escalation with Backfill

**BOIN with backfill (BF-BOIN)**

In this section, we briefly review the BF-BOIN design proposed by Zhao et al. (2023),[25] which combines BOIN with backfill. This review is not only for the completeness of the method but also to provide the necessary notation and concepts for the development of BLRM with backfill (BF-BLRM) in the next section.

BF-BOIN uses the same rule as BOIN for dose escalation and de-escalation. Let $\phi$ denote the target dose-limiting toxicity (DLT) rate, and $\lambda_e$ and $\lambda_d$ denote the corresponding dose escalation and de-escalation boundaries of BOIN, respectively. Let $j = 1, \ldots, J$ denote the dose levels. Let $y_{T_j}$ and $n_j$ denote the number of patients having DLT and the number of patients completed DLT assessment at dose level $j$, respectively, with $\hat{p}_j = y_{T_j}/n_j$ denoting the observed DLT rate at dose $j$. Let $c$ denote the current dose level of dose escalation, where the most recent cohort of patients were treated. The dose escalation/de-escalation of BF-BOIN is given by:

- If $\hat{p}_c \leq \lambda_e$, escalate one dose level to $c + 1$.
- If $\hat{p}_c > \lambda_d$, de-escalate one dose level to $c - 1$.
- Otherwise, stay at the current dose $c$.

For safety, an overdose control rule is applied throughout the dose escalation: if $\Pr(p_j > \phi \mid y_{T_j}, n_j) > C$, dose level $j$ and higher doses are eliminated from treating further patients. This rule is evaluated based on a beta-binomial model with a uniform prior for $p_j$, resulting in the posterior $p_j \mid y_j, n_j \sim Beta(y_j + 1, n_j - y_j + 1)$, where $Beta(\cdot)$ is a beta distribution. The default value of $C$ is 0.95 for BOIN, but it can be adjusted to fit specific trial safety considerations.

To conduct backfilling, BF-BOIN adaptively opens and closes a dose for backfilling based on observed interim data as follows. A dose level $b$ is regarded as eligible for backfilling if it satisfies the following two conditions:

- (Safety) $b$ is lower than the current dose of the dose escalation (i.e., $b < c$).
- (Activity) At least one response is observed at $b$ or at a dose lower than $b$.

Dose level $b$ will be closed for backfilling if:

(a) both of the following two conditions are met:

- The observed DLT rate based on all cumulative patients completed DLT assessment at $b$ is greater than the de-escalation boundary $\lambda_d$, and
- the pooled DLT rate based on the pooled DLT data over $b$ and $b+1$ is also greater than $\lambda_d$. Or,

(b) the number of evaluable patients treated at $b$ is $\geq n_{cap}$, where $n_{cap}$ is a prespecified sample size cap.

The closing rule (a) temporarily halts a dose for backfilling due to its toxicity, while rule (b) closes a dose for backfilling. As described in Zhao et al. (2023),[25] in rule (a), the pooled DLT estimate is used to mitigate the impact of an accidentally high DLT rate caused by a small sample size. This approach is simple and performs similarly to isotonic regression, which is theoretically more desirable but more complex. When multiple doses are eligible for backfilling, we can either prioritize the highest dose or randomize among them,[25] depending on the context. Pin et al. (2024)[26] considered using response-adaptive randomization.

One complication caused by backfilling is that the new data observed from backfilling patients may conflict with those from dose escalation. Specifically, the observed DLT rate at a lower, backfilled dose $b$ ($b < c$) could be higher than the current dose $c$ (for dose escalation), i.e., $\hat{p}_b > \hat{p}_c$. Table 1 provides the possible conflicts between dose escalation and backfilling.

BF-BOIN reconciles the conflict using the following rule. Let $b^*$ denote the dose conflicting the current dose $c$ ($b^* < c$), and define the pooled DLT rate from $b^*$ to $j$, where $b^* \leq j \leq c$, as follows:

$$\hat{q}_j = \frac{\text{the sum of number of patients experienced DLT from dose } b^* \text{ to dose } j}{\text{the sum of number of patients finished DLT assessment from dose } b^* \text{ to dose } j}.$$

In the presence of the conflict, BF-BOIN uses the following rule to replace the original BOIN rule to determine dose escalation/de-escalation:

- If $\hat{q}_c \leq \lambda_e$, escalate one dose level.
- If $\hat{q}_c > \lambda_d$, de-escalate to the highest dose $j$ with $\hat{q}_j \leq \lambda_d$, $b^* \leq j \leq c - 1$; if such a dose does not exist, de-escalate to dose $b^* - 1$.
- Otherwise, stay at the current dose.

BF-BOIN assigns newly enrolled patients to dose escalation or backfilling in the following way:

- If the current cohort of the dose-escalation has not been filled, the patient will be allocated to that dose-escalation cohort;
- Otherwise, the patient will be allocated to a dose that is open for backfilling. If multiple dose levels are open for backfilling, the patient will be assigned to the highest one.

This patient assignment rule prioritizes dose escalation over backfilling, but can be customized based on the trial. Patient enrollment is staggered between cohorts in dose-escalation, and no stagger is necessary in backfilling.

When dose escalation ends (e.g., the prespecified maximum sample size is reached or another stopping rule is satisfied), backfilling stops, and stage 1 ends. At the end of stage 1, an isotonic regression is applied using all the data, and the dose whose isotonic estimated DLT rate is closest to $\phi$ is identified as the MTD.

**BLRM with Backfill (BF-BLRM)**

In this section, we present how to incorporate backfill into BLRM,[22] a model-based design. For convenience, we refer to the resulting design as BF-BLRM. The proposed method is directly applicable to other model-based designs, such as CRM and its extensions.

In BF-BLRM, a Bayesian logistic regression model is used to model the dose-toxicity curve. Let $d_j$ denote the dosage of dose level $j$, and $d^*$ denote a reference dosage. BF-BLRM assumes:

$$\text{logit}(p_j) = \log(\alpha) + \beta\left(\frac{d_j}{d^*}\right), \quad \alpha, \beta > 0.$$

$$\begin{bmatrix}\log(\alpha) \\ \log(\beta)\end{bmatrix} \sim N\left(\begin{bmatrix}\mu_\alpha \\ \mu_\beta\end{bmatrix}, \begin{bmatrix}\sigma_\alpha^2 & 0 \\ 0 & \sigma_b^2\end{bmatrix}\right).$$

To conduct dose escalation/de-escalation, BF-BLRM specifies an underdose cutoff $\gamma_1$ and an overdose cutoff $\gamma_2$, dividing the toxicity probability into three intervals: the underdose interval $[0, \gamma_1]$, the target toxicity interval $(\gamma_1, \gamma_2)$ and the overdose interval $[\gamma_2, 1]$.

Given the observed interim data, BF-BLRM estimates the posterior probability of the target toxicity (PTT) and posterior probability of overdose (POD) as $PTT_j = \Pr(p_j \in (\gamma_1, \gamma_2)|data)$ and $POD_j = \Pr(p_j \in [\gamma_2, 1]|data), j = 1, \cdots, J$. BF-BLRM identifies the dose $j^*$ that has the highest value of PTT with POD $< \eta$ among $J$ doses, where $\eta$ is a prespecified overdose control cutoff. Following the escalation with overdose control method,[27] the authors of BLRM recommended $\eta = 0.25$. However, this recommendation was found to be overly conservative and led to poor accuracy in selecting the MTD.[28-31] In BF-BLRM, we use $\eta = 0.30$ to improve its operating characteristics. BF-BLRM conducts dose-escalation/de-escalation as follows:

- If $j^* > c$, escalate one dose level to $c + 1$.
- If $j^* < c$, de-escalate one dose level to $c - 1$.
- Otherwise, stay at the current dose $c$.

At any time of the trial, if the POD $> \eta$ for all doses, we terminate the trial and claim that all doses are over-toxic. In this case, no dose should be selected as the MTD.

BF-BLRM incorporates backfilling into dose escalation similarly to BF-BOIN. Specifically, BF-BLRM adaptively opens and closes a dose level $b$ for backfilling using criteria similar to BF-BOIN, with a modification to the rule for closing a dose:

A dose level $b$ will be closed for backfilling if:

(a) POD for dose $b$ is $\geq \eta$.

(b) The number of evaluable patients treated at $b$ is $\geq n_{cap}$.

Patients are assigned to the current dose-escalation cohort or backfilled using the same approach as in BF-BOIN.

BF-BLRM assumes a dose-toxicity model that incorporates data across all doses and imposes monotonicity by assuming $\beta > 0$, Therefore, data conflicts between the backfilled dose $b$ and the current dose $c$ are automatically reconciled and smoothed out by fitting the model, obviating the need for additional rules.

At the end of stage 1, BF-BLRM selects the MTD based on all data, including dose escalation and backfilling data. The MTD is chosen as the dose that satisfies the following two conditions:

- Treated with at least 6 patients,
- Has the highest PTT among all doses with POD $< \eta$.

## Stage 2 with Adaptive Randomization

Suppose at the end of stage 1, two doses are selected for stage 2 randomization, referred to as $d_{low}$ and $d_{high}$ with $d_{low} < d_{high}$. This selection is based on the evaluation of the totality of stage 1 data, including safety, efficacy, PK/PD, and tolerability. Often, $d_{high}$ is the MTD. For ease of exposition, we assume two doses, but the method is directly applicable to more than two doses.

To optimize the dose, the most straightforward approach is to randomize patients to $d_{low}$ and $d_{high}$ in a fixed ratio, most commonly 1:1. Randomization is preferred, as noted in FDA's guidance,[16] because it balances important prognostic factors between the two doses, allowing an unbiased comparison of their risk-benefit profiles to select the OBD. Since some patients were treated with $d_{low}$ and $d_{high}$ in stage 1, it is highly desirable to incorporate this stage 1 data with stage 2 data to enhance trial efficiency and reduce the sample size for dose optimization. The challenge is that incorporating non-randomized stage 1 data with randomized stage 2 data may compromise the balance of the latter, defeating the purpose of stage 2 randomization.

To address this challenge, we leverage the idea of the Pocock-Simon minimization[32] for stage 2 randomization. Minimization is a widely used covariate-adaptive randomization method that is discussed in the FDA's guidance on adaptive designs.[33] Our key idea is to randomize stage 2 patients, conditional on stage 1 data, in a covariate-adaptive way to eliminate the covariate imbalance present in the stage 1 data. By doing so, at the end of stage 2, covariates are balanced between the two dose arms. This allows stage 1 and stage 2 data to be combined to better inform dose comparison and the selection of OBD.

Let $n_{1,low}$ and $n_{1,high}$ denote the number of patients treated at dose $d_{low}$ and $d_{high}$, respectively, in stage 1. In cases where patient enrollment criteria differ between stage 1 and stage 2 (e.g., stage 1 enrolls all-comers and stage 2 enrolls specific indications), $n_{1,low}$ and $n_{1,high}$ refer to subsets of patients who meet the stage 2 eligibility criteria. Let $N_2$ denote the target total sample size to be treated with $d_{low}$ and $d_{high}$ by the end of the trial, including those from both stage 1 and 2. Then, $N_2^* = N_2 - n_{1,low} - n_{1,high}$ new patients will be enrolled and randomized in stage 2. Yang et al. (2023)[20] proposed a method to determine $N_2$ and recommended that a sample size of 20 to 40 patients per dose is often reasonable for randomized dose optimization.

Let $X_1$ and $X_2$ denote important baseline prognostic factors that we aim to balance via randomization. For illustrative purposes, we consider two prognostic factors, but the method is applicable to more than two factors. We assume that $X_1$ and $X_2$ are categorical with $L_1$ and $L_2$ levels, respectively. When prognostic factors are continuous, they can be discretized to balance their distribution between the two dose arms. Let $n_{low}(X_k = l)$ be the number of patients with $X_k = l$ who are treated with $d_{low}$, and $n_{high}(X_k = l)$ is the number of patients with $X_k = l$ who are treated with $d_{high}$, where $k = 1, 2$. The difference $|n_{low}(X_k = l) - n_{high}(X_k = l)|$ provides a measure of imbalance on level $l$ of $X_k$ between $d_{low}$ and $d_{high}$.

When a new patient with $X_1 = v_1$ and $X_2 = v_2$ is enrolled at stage 2, the dose assignment of this patient only impacts the balance of $X_1$ at level $v_1$ and $X_2$ at level $v_2$. Following Pocock and Simon,[31] define the imbalance index that embraces both $X_1$ and $X_2$ as:

$$\omega = \sum_{k=1}^{2} |n_{low}(X_k = v_k) - n_{high}(X_k = v_k)|. \tag{1}$$

To balance the distribution $X_1$ and $X_2$ between two dose arms, the new patient will be assigned with probability $r$ to the dose ($d_{low}$ or $d_{high}$) that minimizes $\omega$, where $r \leq 1$ is a large probability between 0.8 to 1. When $r = 1$, the patient is always assigned to the dose that minimizes $\omega$. FDA's Guidance for adaptive designs[33] noted that setting $r < 1$ reduces the predictability of treatment assignment.

It is important to note that the calculation of $n_{low}(X_k = v_k)$ and $n_{high}(X_k = v_k)$ in (1) is based on both stage 1 and 2 data. Thus, our approach can be viewed as a conditional version of the Pocock-Simon method, where the randomization of stage 2 patients is conditional on stage 1 data, to handle the mixture of non-randomized and randomized patients. By doing so, we actively rebalance prognostic factors that may not be well balanced in stage 1. Thus, at the end of randomization, the combined data of stage 1 and stage 2 efficiently resembles these generated by randomization of full $N_2$ patients. In contrast, Pocock-Simon method focuses on "full" randomization and the assignment of the next patient based on the covariate distribution of patients who have been randomized. Because substantial imbalance might be present in the stage 1 data and the sample size of stage 2 is often small, we generally recommend using a large value of $r$ (e.g., 0.95 or 0.9) to quickly eliminate the imbalance.

## OBD Selection

At the end of stage 2, we identify the OBD based on data from $N_2$ patients, combined from stages 1 and 2. Depending on the trial setting, different criteria can be used to select the OBD. We consider two approaches, noting that their modifications and other criteria can also be used to define and select the OBD. Let $\hat{p}_{E,low}$ and $\hat{p}_{E,high}$ denote the estimates of the efficacy rate for $d_{low}$ and $d_{high}$, respectively. These estimates can be sample mean (e.g., observed efficacy rates) or estimates based on a certain model (e.g., logistic model).

The first approach implicitly considers the toxicity-efficacy trade-off and select the OBD as follows:

- If $\hat{p}_{E,low} - \hat{p}_{E,high} \geq \delta$, select $d_{low}$ as the OBD; otherwise, select $d_{high}$ as the OBD.

where $\delta$ is a prespecified noninferiority/indifference margin. The rationale behind this criterion is that $d_{low}$ is presumably safer than $d_{high}$. Therefore, if the efficacy of $d_{low}$ is noninferior to $d_{high}$, $d_{low}$ has a better toxicity-efficacy trade-off and should be selected as the OBD.

The second method explicitly accounts for the toxicity-efficacy trade-off and selects the OBD based on utility. For binary toxicity and efficacy endpoints, each patient can experience one of four possible outcomes: (toxicity, no efficacy), (no toxicity, no efficacy), (toxicity, efficacy), and (no toxicity, efficacy). Let $u_1, \cdots, u_4$ denote the utility scores assigned to these outcomes, which should be elicited from clinicians to reflect the relative desirability of each outcome. Typically, $u_1$ is assigned a score of 0 (least desirable, toxicity without efficacy), and $u_4$ a score of 100 (most desirable, no toxicity with efficacy). Clinicians then provide scores for the other outcomes based on this reference. Table 2 provides an example of elicited utility scores.

Let $\pi_{j1}, \pi_{j2}, \pi_{j3}$ and $\pi_{j4}$ denote the probabilities of occurrence of these four outcomes at dose level $j$. We assume the prior distribution of $\pi_j = (\pi_{j1}, \pi_{j2}, \pi_{j3}, \pi_{j4})$ is Dirichlet($\alpha_1, \alpha_2, \alpha_3, \alpha_4$). Let $n_{j1}, n_{j2}, n_{j3}$, and $n_{j4}$ denote the numbers of patients of these four outcomes who were treated at dose level $j$. Applying Dirichlet-multinomial model, the posterior distribution of $\pi_j$ is:

$$\pi_j|\text{data} \sim \text{Dirichlet}\left(\alpha_1 + n_{j1}, \alpha_2 + n_{j2}, \alpha_3 + n_{j3}, \alpha_4 + n_{j4}\right).$$

The posterior mean utility of dose $j$ is estimated as:

$$\widehat{U}_j = \sum_{k=1}^{4} u_k E(\pi_{jk}|\text{data}).$$

$d_{low}$ or $d_{high}$ that maximize $\widehat{U}_j$ is selected as OBD.

In both approaches, we require that the OBD $j$ also satisfies the following safety and efficacy requirements:

- (Safety) $\Pr(p_j > \phi_T|\text{data}) \leq C_T,$
- (Efficacy) $\Pr(p_{E,j} < \phi_E|\text{data}) \leq C_E,$

where $\phi_T$ and $\phi_E$ are the upper and lower limits of the toxicity and efficacy rates, respectively, and $C_T$ and $C_E$ are probability cutoffs calibrated through simulation. Typically, $C_T$ and $C_E$ are set to relatively high values, such as 0.8 to 0.95, to minimize the probability of incorrectly ruling out safe and effective doses. In the case that $d_{low}$ had a higher posterior probability of over-toxic than $d_{high}$, isotonic-adjusted posterior probabilities were applied to evaluate the safety condition. If only one of $d_{low}$ and $d_{high}$ satisfies the safety and efficacy requirements, that dose will be selected as the OBD.

To facilitate the use of the BARD design, software is available at www.trialdesign.org, allowing users to run simulations and conduct trials.

# Numerical studies

## Simulation setting

We considered a trial where stage 1 aims to find the MTD from 5 doses with a maximum sample size of 30 for dose escalation. The dose escalation starts from the lowest dose, and patients are treated in cohorts of 3. The accrual rate is 3/month, and the DLT assessment window is 1 month. The sample size cap for backfilling is $n_{cap}$=12 per dose. At the end of stage 1, the identified MTD and the dose one level lower (if it exists) move forward to stage 2 for randomization. The targeted total sample size for stage 2 is $N_2 = 40$, with 20 patients per dose arm. The randomization parameter $r = 0.95$ is used in stage 2 to assign patients to the arm that minimizes covariate imbalance.

We compared two BARD designs, BARD-BOIN and BARD-BLRM, with their conventional counterparts, referred to as BOIN-SR and BLRM-SR, where BOIN or BLRM is used for stage 1 dose escalation followed by 1:1 simple randomization. In BOIN-SR and BLRM-SR, stage 1 data are not combined with stage 2 data. Thus, a total of 40 new patients are enrolled and randomized in stage 2 to reach 20 patients per dose arm.

For BARD-BOIN and BOIN-SR, the target DLT rate is set at $\phi = 0.25$, with the corresponding escalation and de-escalation boundaries being $\lambda_e = 0.197$ and $\lambda_d = 0.298$, respectively. Stage 1 dose escalation ends when the maximum dose-escalation sample size of 30 is reached, or the number of patients treated at the current dose reaches $n_{stop} = 9$ and the decision is "stay". The default cutoff $C = 0.95$ is used in its overdose control rule.

For BARD-BLRM and BLRM-SR, the target toxicity interval is set at $(0.16, 0.33)$. The dosages are set as 10, 20, 50, 100, 200 with the reference dosage as 50. The following weakly-informative prior suggested by Neuenschwander et al (2015)[34] is used to fit the model:

$$\begin{bmatrix} \log(\alpha) \\ \log(\beta) \end{bmatrix} \sim N\left(\mu = \begin{bmatrix} -1.1 \\ 0 \end{bmatrix}, V = \begin{bmatrix} 4 & 0 \\ 0 & 1 \end{bmatrix}\right).$$

As described previously, the EWOC cutoff is set at $\eta = 0.30$. Unlike BOIN, BLRM does not have a rule for stopping the trial when the number of patients treated at the current dose reaches $n_{stop} = 9$ and the decision is "stay". To facilitate comparison, we calibrated the maximum stage 1 dose-escalation sample size of BARD-BLRM and BLRM-SR so that their average sample size in stage 1 matches that of BARD-BOIN and BOIN-SR.

To evaluate the performance of BARD-BOIN and BARD-BLRM in balancing covariates, we assumed three binary prognostic factors $X_1, X_2$ and $X_3$, generated from Bernoulli (0.5), which are related to the efficacy rate $p_E$ as follows: $\text{logit}(p_E|d_j) = \beta_{0j} + \beta_1 X_1 + \beta_2 X_2 + \beta_3 X_3$, where $\beta_1 = 1.7, \beta_2 = -1.5, \beta_3 = 0.4$ and the values of intercepts $\beta_{0j}$ can be found in Supplementary Table S1. In our stage 2 covariate-adaptive randomization algorithm, we included only $X_1$ and $X_2$. We intentionally omitted $X_3$ from our algorithm to compare the balance of covariates when it is included versus when it is not included. In addition, we included Simon-Pocock randomization with 40 patients as the benchmark for comparison.

At the end of stage 2, the OBD is identified by the two approaches described previously. In the first efficacy-rate-based approach, we used $\delta = 0.05$ to select the OBD. In the utility-based method, we assigned utility scores according to Table 2. We set $\phi_T = 0.3$, $\phi_E = 0.2$, $C_T = 0.9$, and $C_E = 0.95$ to ensure safety and efficacy of the OBD.

We considered 8 representative scenarios that differ in the toxicity-response curves and the location of the OBD, as presented in Table 3. In scenarios 1-4, the toxicity-response curve is monotone increasing, while in scenarios 5-8, the response rate plateaus below the MTD. The following performance metrics were evaluated based on 30,000 simulated trials.

- Average total sample size.
- Average trial duration.
- Imbalance index, defined as the absolute difference between the proportion of patients with $X_k = 1$ in $d_{low}$ and $d_{high}$, $k = 1, 2, 3$, which measures the imbalance of the distribution of $X_k$ between $d_{low}$ and $d_{high}$. A smaller value of imbalance index indicates a better balance.
- Imbalance in allocation, defined as the absolute difference in the number of patients treated at $d_{low}$ and that treated at $d_{high}$.
- PCS1: the percentage of correct selection (PCS) of the true OBD based on the efficacy-rate-based approach.
- PCS2: the PCS of the true OBD based on the utility approach.

# Results

Table 4 summarizes the operating characteristics of the designs. BARD-BOIN outperforms its counterpart, BOIN-SR. Across all eight scenarios, BARD-BOIN reduces the sample size by 10-15 patients and shortens the trial duration by 6-8 months compared to BOIN-SR, because of the integration of stage 1 and stage 2 data. Additionally, BARD-BOIN achieves significantly better balance on $X_1$ and $X_2$ than BOIN-SR, with the imbalance index for BARD-BOIN often being one-third that of BOIN-SR. This highlights the effectiveness of the proposed covariate-adaptive randomization, which yields superior covariate balance compared to the "full" simple randomization (without combining stage 1 data). The covariate balance under BARD-BOIN is very similar to that of the "full" Pocock-Simon assignment (without combining stage 1 data), further confirming the approach's effectiveness. For the omitted covariate $X_3$, the balance remains comparable to that of full simple randomization, indicating that the presence of unknown prognostic factors does not compromise the proposed method. The accumulative number of patients allocated to $d_{low}$ and $d_{high}$ is nearly 1:1, with the average difference generally less than one patient.

In terms of OBD selection, BARD-BOIN generally outperforms BOIN-SR, with a 1.74% higher PCS1 and 1.45% higher PCS2 on average. For example, in scenarios 6 and 7, the PCS1 of BARD-BOIN is 3.52% and 3.23% percentage points higher than that of BOIN-SR. This result is remarkable, considering that BARD-BOIN uses a smaller sample size.

As for the two OBD selection approaches (i.e., PCS1 and PCS2), they are generally comparable. Since they are based on different criteria, reflecting distinct clinical considerations and suited to different clinical settings, it is more meaningful to focus on their overall operating characteristics rather than making direct comparisons and drawing general conclusions about which approach is superior.

Similar patterns are observed when comparing BARD-BLRM to BLRM-SR across these performance metrics. Specifically, BARD-BLRM reduces the sample size by 12-15 patients and shortens the trial duration by 6-8 months compared to BLRM-SR. Additionally, BARD-BLRM demonstrates greater accuracy in identifying the OBD, with higher PCS, and a superior ability to balance covariates compared to BLRM-SR.

Between BARD-BOIN and BARD-BLRM, BARD-BOIN often exhibits higher accuracy in identifying the OBD, as evidenced by higher PCS1 and PCS2. This is primarily due to BLRM's lower probability of correctly identifying the MTD. Table S2 in the Supplementary Materials summarizes stage 1 of the BARD-BOIN and BARD-BLRM designs in the simulation. BARD-BLRM has a lower probability for carrying forward the true OBD dose to stage 2. Our results align with previous findings that BLRM tends to be overly conservative, resulting in a lower probability of identifying the MTD.[28-31]

It was somewhat unexpected that BARD-BLRM showed notably worse covariate balance than BARD-BOIN, although it still outperformed the simple randomization. This result is surprising, given that both designs use the same covariate-adaptive randomization method in stage 2. A key factor contributing to this result is the rigidity of BLRM due to the use of the two-parameter logistic model. The concept of "rigidity" is defined and discussed in Cheung et al. (2011)[35] and Iasonos et al. (2016).[36] It refers to the tendency of a flexible model to overfit the data, which in turn causes the dose-finding process to become stuck at a low dose, preventing exploring higher doses that seem toxic based on the data from a few patients (e.g., 3), which are actually safe. Once the process is stuck at a dose, treating more patients does not resolve the issue. Given the limited data available at the beginning of a dose-finding trial (e.g., data from only 3 or 6 patients), the two-parameter logistic model is often deemed overly flexible, leading to overfitting and getting stuck at a particular dose. As a result, BLRM often leads to a highly imbalanced number of patients between $d_{low}$ and $d_{high}$ at the end of stage 1. This imbalance is carried over to stage 2, making it challenging to fully correct given the limited sample size in that stage. Supplementary Materials Section 3 provide further explanation and numerical results on this issue.

## Sensitivity analysis

We conducted a sensitivity analysis to assess the robustness of BARD-BOIN regarding the number of covariates, the stage 2 sample size $N_2$, as well as the number of doses $J$. We focused on BARD-BOIN due to its superior performance in balancing covariates. Figure 1 depicts the difference in the covariate imbalance index for $X_1$ between BARD-BOIN and "full" Simon-Pocock randomization with 2, 3, and 4 covariates when $N_2 = 40$, as well as when $N_2$ is 40, 60, and 80, given 2 covariates. The imbalance in BARD-BOIN is close to that of full Simon-Pocock randomization, with no more than a 1.5% higher imbalance index in general. Due to the symmetric role of covariates, the covariate imbalance index for the other covariates is similar to that of $X_1$; therefore, only $X_1$ is displayed. Tables S4-6 in Supplementary Materials show the result with $J = 3$ doses. The results are generally consistent with those observed with 5 doses. Specifically, compared to BOIN-SR, BARD-BOIN reduces the sample size and trial duration, achieves better covariate balance, and improves the accuracy of identifying the OBD.

## Discussion

We have proposed a seamless two-stage design, BARD, that integrates backfilling and adaptive randomization for efficient dose optimization. Backfilling allows additional patients to be enrolled at doses deemed safe and showing promising activity, enhancing patient enrollment and data generation without extending the trial duration. The adaptive randomization enables the combination of data from dose escalation and randomization without compromising the balance of baseline characteristics between comparative dose arms. BARD designs offer an efficient solution to meet the dose optimization requirements set by Project Optimus.

Backfilling and adaptive randomization significantly enhance trial efficiency when used together, but they do not necessarily need to be bundled. For instance, if efficacy data cannot be observed quickly enough to allow for efficient backfilling, stage 1 dose escalation can proceed without backfilling, while still utilizing adaptive randomization to combine stage 1 and 2 data for a more efficient comparison of multiple doses. Additionally, while we focus on using the Pocock-Simon method, its various extensions and other covariate-balance randomization[37] methods can also be employed when appropriate.

Because dose selection is performed at the end of stage 1, one may concern potential estimation bias when stage 1 and 2 data are combined to estimate and select the OBD at the end of the trial, an issue analogous to what occurs in inferential phase II-III trials.[38] Table S7 presents the bias of the estimates of the DLT rate and response rate at the end of the trial. The estimate of the response rate has minimal bias. The estimate of the DLT rate has small negative bias (-0.015 to -0.021), but is generally negligible compared to the high heterogeneity of typical early phase patients and the variance of the DLT estimate.

The stage 1 of BARD designs centers on dose escalation based on DLT and the identification of MTD. When suitable, efficacy-integrated designs, such as EffTox and BOIN12, can be used to more efficiently identify doses likely to be the OBD, which can then be advanced to stage 2 for adaptive randomization. Additionally, our simulation does not include interim toxicity and futility monitoring, which potentially further reduces the sample size if one or two doses in stage 2 are overly toxic or futile. Bayesian optimal phase 2 design[39, 40] can be employed to achieve this goal. Finally, this paper focuses on single-agent dose-finding trials. Extending BARD to combination trials involving the identification of the OBD from a dose matrix is a topic for future research.

# References

1. Ratain M. Redefining the primary objective of phase I oncology trials. *Nature Rev Clin Oncol* 2014; 11: 503–504.

2. Yan F, Thall P and Yuan Y. Phase I-II clinical trial design: a state-of-the-art paradigm for dose finding. *Ann Oncol* 2018; 29: 694–699.

3. Shah M, Rahman A, Theoret MR, et al. The Drug-Dosing Conundrum in Oncology — When Less Is More. *N Engl J Med* 2021; *385*(16): 1445–1447.

4. U.S. Food & Drug Administration. Project Optimus: Reforming the dose optimization and dose selection paradigm in oncology, https://www.fda.gov/about-fda/oncology-center-excellence/project-optimus. (2024, accessed 13 July 2024).

5. Yuan Y, Nguyen H and Thall P. *Bayesian designs for phase I-II clinical trials.* 1st ed. New York: Chapman and Hall/CRC, 2016.

6. Yuan Y, Zhou H and Liu S. Statistical and practical considerations in planning and conduct of dose-optimization trials. *Clin Trials* 2024; 21(3): 273–286.

7. Thall PF and Cook JD. Dose-finding based on efficacy toxicity trade-offs. *Biometrics* 2004; 60(3): 684–693.

8. Jin IH, Liu S, Thall PF, et al. Using Data Augmentation to Facilitate Conduct of Phase I-II Clinical Trials with Delayed Outcomes. *J Am Stat Assoc* 2014; 109(506): 525–536.

9. Lin R, Zhou Y, Yan F et al. BOIN12: Bayesian Optimal Interval Phase I/II Trial Design for Utility-Based Dose Finding in Immunotherapy and Targeted Therapies. *JCO Precis Oncol* 2020; *4*(4): 393–1402.

10. Takeda K, Taguri M and Morita S. BOIN-ET: Bayesian optimal interval design for dose finding based on both efficacy and toxicity outcomes. *Pharm Stat* 2018;17(4): 383-395.

11. Shi H, Cao J, Yuan Y, et al. uTPI: A utility-based toxicity probability interval design for phase I/II dose-finding trials. *Stat Med* 2021; 40(11): 2626-2649.

12. National Library of Medicine. Intratumoral injections of LL37 for melanoma, https://clinicaltrials.gov/ct2/show/record/NCT02225366. (2021, accessed 13 July 2024).

13. Msaouel P, Goswami S, Thall PF, et al. A phase 1-2 trial sitravatinib and nivolumab in clear cell renal cell carcinoma following progression on antiangiogenic therapy. *Sci Trans Med* 2022; 14(641): eabm6420.

14. Pan J, Tan Y, Shan L, et al. Phase I study of donor derived CD5 CAR T cells in patients with relapsed or refractory T-cell acute lymphoblastic leukemia. *J Clin Oncol* 2022; 4: 7028–7028.

15. Simon R, Wittes RE, and Ellenberg SS. Randomized phase II clinical trials. *Cancer Treat Rep* 1985; 69(12): 1375.

16. U.S. Food & Drug Administration. Optimizing the Dosage of Human Prescription Drugs and Biological Products for the Treatment of Oncologic Diseases, https://www.fda.gov/regulatory-information/search-fda-guidance-documents/optimizing-dosage-human-prescription-drugs-and-biological-products-treatment-oncologic-diseases. (2021, accessed 13 July 2024).

17. Hoering A, LeBlanc M and Crowley J. Seamless phase I-II trial design for assessing toxicity and efficacy for targeted agents. *Clin Can Res* 2011; 17(4): 640–646.

18. Guo B. and Yuan Y. Droid: Dose-ranging approach to optimizing dose in oncology drug development. *Biometrics* 2023; 79(4): 2907–2919.

19. Zhou, Y, Lee JJ and Yuan Y. A utility-based Bayesian optimal interval (U-BOIN) phase I/II design to identify the optimal biological dose for targeted and immune therapies. *Stat Med* 2019; 38(28): 5299–5316.

20. Yang P, Li D, Lin R, et al. Design and sample size determination for multiple-dose randomized phase II trials for dose optimization. arXiv Preprint, https://arxiv.org/abs/302.09612. (2023, accessed 13 July 2024).

21. Yuan Y, Hess KR, Hilsenbeck SG, et al. Bayesian Optimal Interval Design: A Simple and Well-Performing Design for Phase I Oncology Trials. *Clin Cancer Res* 2016; 22(17): 4291–4301.

22. Neuenschwander B, Branson M and Gsponer T. Critical aspects of the Bayesian approach to phase I cancer trials. *Stat Med* 2008; 27(13): 2420-2439.

23. Yan F, Mandrekar SJ and Yuan Y. Keyboard: A novel Bayesian toxicity probability interval design for phase I clinical trials. *Clin Cancer Res*; 23(15): 3994–4003.

24. O'Quigley J, Pepe M, and Fisher L. Continual Reassessment Method: A Practical Design for Phase 1 Clinical Trials in Cancer. *Biometrics* 1990; 46 (1): 33–48.

25. Zhao Y, Yuan Y, Korn E, et al. Backfilling Patients in Phase I Dose-Escalation Trials Using Bayesian Optimal Interval Design (BOIN). *Clin Cancer Res* 2024; 30(4): 673-OF7.

26. Pin L, Villar SS and Dehbi HM. Implementing and assessing Bayesian response-adaptive randomisation for backfilling in dose-finding trials. *Contemp Clin Trials* 2024; 142:107567. doi:10.1016/j.cct.2024.107567.

27. Babb J, Rogatko A and Zacks S. Cancer phase I clinical trials: efficient dose escalation with overdose control. *Stat Med* 1998; 17(10): 1103–1120.

28. Zhou H, Yuan Y and Nie L. Accuracy, Safety, and Reliability of Novel Phase I Trial Designs. *Clin Cancer Res* 2018; 24(18): 4357–4364.

29. Zhang H, Chiang AY, Wang J. Improving the performance of Bayesian logistic regression model with overdose control in oncology dose finding studies. *Stat Med* 2022; 41:5463-5483.

30. Yuan Y and Zhao Y. Commentary on "Improving the performance of Bayesian logistic regression model with overdose control in oncology dose-finding studies". *Stat Med* 2022; 41(27): 5484–5490.

31. Yang CH, Cheng G and Lin R. On the relative conservativeness of Bayesian logistic regression method in oncology dose-finding studies. *Pharm Stat* 2024; 23(4): 585-594.

32. Pocock SJ and Simon R. Sequential Treatment Assignment with Balancing for Prognostic Factors in the Controlled Clinical Trial. *Biometrics* 1975; 31(1): 103–115.

33. U.S. Food & Drug Administration. Adaptive Designs for Clinical Trials of Drugs and Biologics Guidance for Industry, https://www.fda.gov/regulatory-information/search-fda-guidance-documents/adaptive-design-clinical-trials-drugs-and-biologics-guidance-industry.(2019, accessed 28 Aug 2024)

34. Neuenschwander B, Matano A, Tang Z and et al. A Bayesian Industry Approach to Phase I Combination Trials in Oncology. In: Zhao W and Yang H *Statistical Methods in Drug Combination Studie.* 1st ed. New York: Chapman and Hall/CRC, 2015, pp. 112–153.

35. Cheung YK. *Dose finding by the continual reassessment method.* 1st ed. New York: Chapman and Hall/CRC, 2011.

26
25. Zhao Y, Yuan Y, Korn E, et al. Backfilling Patients in Phase I Dose-Escalation Trials Using Bayesian Optimal Interval Design (BOIN). *Clin Cancer Res* 2024; 30(4): 673-OF7.

26. Pin L, Villar SS and Dehbi HM. Implementing and assessing Bayesian response-adaptive randomisation for backfilling in dose-finding trials. *Contemp Clin Trials* 2024; 142:107567. doi:10.1016/j.cct.2024.107567.

27. Babb J, Rogatko A and Zacks S. Cancer phase I clinical trials: efficient dose escalation with overdose control. *Stat Med* 1998; 17(10): 1103–1120.

28. Zhou H, Yuan Y and Nie L. Accuracy, Safety, and Reliability of Novel Phase I Trial Designs. *Clin Cancer Res* 2018; 24(18): 4357–4364.

29. Zhang H, Chiang AY, Wang J. Improving the performance of Bayesian logistic regression model with overdose control in oncology dose finding studies. *Stat Med* 2022; 41:5463-5483.

30. Yuan Y and Zhao Y. Commentary on "Improving the performance of Bayesian logistic regression model with overdose control in oncology dose-finding studies". *Stat Med* 2022; 41(27): 5484–5490.

31. Yang CH, Cheng G and Lin R. On the relative conservativeness of Bayesian logistic regression method in oncology dose-finding studies. *Pharm Stat* 2024; 23(4): 585-594.

32. Pocock SJ and Simon R. Sequential Treatment Assignment with Balancing for Prognostic Factors in the Controlled Clinical Trial. *Biometrics* 1975; 31(1): 103–115.

33. U.S. Food & Drug Administration. Adaptive Designs for Clinical Trials of Drugs and Biologics Guidance for Industry, https://www.fda.gov/regulatory-information/search-fda-guidance-documents/adaptive-design-clinical-trials-drugs-and-biologics-guidance-industry.(2019, accessed 28 Aug 2024)

34. Neuenschwander B, Matano A, Tang Z and et al. A Bayesian Industry Approach to Phase I Combination Trials in Oncology. In: Zhao W and Yang H *Statistical Methods in Drug Combination Studie.* 1st ed. New York: Chapman and Hall/CRC, 2015, pp. 112–153.

35. Cheung YK. *Dose finding by the continual reassessment method.* 1st ed. New York: Chapman and Hall/CRC, 2011.

36. Iasonos A, Wages NA, Conaway MR, et al. Dimension of model parameter space and operating characteristics in adaptive dose-finding studies. *Stat Med* 2016; 35(21): 3760–3775.

37. Scott NW, McPherson GC, Ramsay CR, et al. The method of minimization for allocation to clinical trials: a review. *Control Clin Trials* 2002; 23(6): 662–674.

38. Jiang L and Yuan Y. Seamless phase II/III design: a useful strategy to reduce the sample size for dose optimization. *J Natl Cancer Inst* 2023;115(9):1092-1098. doi:10.1093/jnci/djad103

39. Zhou H, Lee JJ, Yuan Y. (2017) BOP2: Bayesian optimal design for phase II clinical trials with simple and complex endpoints. *Stat Med.* 36(21):3302-3314.

40. Chen K, Zhou H, Lee JJ, et al. BOP2-TE: Bayesian optimal phase 2 design for jointly monitoring efficacy and toxicity with application to dose optimization. *J Biopharm Stat* 2024; revision submitted.

Table 1. Conflict between the current dose of dose-escalation and backfilling doses.

| Observed DLT rate of backfilled dose $\hat{p}_b$ suggests | Observed DLT rate of the current dose $\hat{p}_c$ suggests | | |
|---|---|---|---|
| | Escalation ($\hat{p}_c \leq \lambda_e$) | stay ($\lambda_e < \hat{p}_c \leq \lambda_d$) | de-escalation ($\hat{p}_c > \lambda_d$) |
| Escalation ($\hat{p}_b \leq \lambda_e$) | | | |
| Stay ($\lambda_e < \hat{p}_b \leq \lambda_d$) | conflict | | |
| de-escalation ($\hat{p}_b > \lambda_d$) | conflict | conflict | conflict* |

*This case does not necessarily mean that $\hat{p}_b > \hat{p}_c$. However, as $\hat{p}_b > \lambda_d$, it means that the additional data from backfilling patients demonstrate alarmingly higher toxicity than what originally observed during the dose escalation (i.e., $\hat{p}_b \leq \lambda_e$). As $b < c$, the data observed at $b$ previously during the dose escalation must satisfy $\hat{p}_b \leq \lambda_e$, Therefore, it is important to reconcile such conflict for patient safety.

Table 2. Utility ascribed to each possible efficacy-toxicity outcome.

| | Toxicity | No toxicity |
|---|---|---|
| No efficacy | 0 | 30 |
| Efficacy | 50 | 100 |

Table 3. Simulation scenarios, with the OBD highlighted in bold.

| Scenario | | Dose level | | | | |
|---|---|---|---|---|---|---|
| | | 1 | 2 | 3 | 4 | 5 |
| 1 | DLT | 0.12 | **0.25** | 0.42 | 0.49 | 0.55 |
| | Efficacy | 0.181 | **0.349** | 0.439 | 0.519 | 0.596 |
| | Utility | 38.6 | **45.2** | 44.4 | 46.6 | 48.7 |
| 2 | DLT | 0.04 | 0.12 | **0.25** | 0.43 | 0.63 |
| | Efficacy | 0.152 | 0.181 | **0.349** | 0.439 | 0.519 |
| | Utility | 39.3 | 38.6 | **45.2** | 44.0 | 40.9 |
| 3 | DLT | 0.02 | 0.06 | 0.1 | **0.25** | 0.4 |
| | Efficacy | 0.103 | 0.152 | 0.181 | **0.349** | 0.439 |
| | Utility | 36.6 | 38.6 | 39.3 | **45.2** | 45.2 |
| 4 | DLT | 0.02 | 0.05 | 0.08 | 0.11 | **0.25** |
| | Efficacy | 0.046 | 0.103 | 0.152 | 0.181 | **0.349** |
| | Utility | 32.6 | 35.6 | 38.0 | 38.9 | **45.2** |
| 5 | DLT | **0.12** | 0.25 | 0.42 | 0.49 | 0.55 |
| | Efficacy | **0.349** | 0.349 | 0.359 | 0.359 | 0.359 |
| | Utility | **50.0** | 45.2 | 39.5 | 36.9 | 34.7 |
| 6 | DLT | 0.04 | **0.12** | 0.25 | 0.43 | 0.63 |
| | Efficacy | 0.181 | **0.349** | 0.349 | 0.359 | 0.359 |
| | Utility | 41.3 | **50.0** | 45.2 | 39.1 | 31.7 |
| 7 | DLT | 0.02 | 0.06 | **0.1** | 0.25 | 0.4 |
| | Efficacy | 0.152 | 0.181 | **0.349** | 0.349 | 0.359 |
| | Utility | 40.0 | 40.6 | **50.8** | 45.2 | 40.3 |
| 8 | DLT | 0.02 | 0.05 | 0.08 | **0.11** | 0.25 |
| | Efficacy | 0.103 | 0.152 | 0.181 | **0.349** | 0.349 |
| | Utility | 36.6 | 39.0 | 39.9 | **50.4** | 45.2 |

Table 4. Operating characteristics of BARD-BOIN and BARD-BLRM, in comparison with BOIN-SR and BLRM-SR.

| Design | N | Duration (month) | Imbalance* $X_1$ | Imbalance* $X_2$ | Imbalance* $X_3$ | Imbalance* allocation | PCS1 | PCS2 |
|---|---|---|---|---|---|---|---|---|
| Scenario 1 | | | | | | | | |
| BARD-BOIN | 39.37 | 17.75 | 4.50 (3.50) | 4.52 (3.46) | 12.51 (12.64) | 0.99 (0.83) | 51.43 | 48.44 |
| BARD-BLRM | 32.26 | 15.35 | 7.53 (3.49) | 7.54 (3.46) | 12.72 (12.58) | 2.44 (0.83) | 46.93 | 44.16 |
| BOIN-SR | 54.60 | 24.56 | 12.61 | 12.55 | 12.55 | 0 | 50.10 | 48.42 |
| BLRM-SR | 46.45 | 21.39 | 12.49 | 12.50 | 12.37 | 0 | 43.02 | 41.61 |
| Scenario 2 | | | | | | | | |
| BARD-BOIN | 48.62 | 20.87 | 4.16 (3.51) | 4.12 (3.43) | 12.61 (12.59) | 0.87 (0.83) | 49.79 | 47.55 |
| BARD-BLRM | 44.14 | 19.73 | 6.33 (3.52) | 6.32 (3.49) | 12.66 (12.52) | 1.85 (0.82) | 31.39 | 30.60 |
| BOIN-SR | 63.30 | 28.42 | 12.58 | 12.53 | 12.54 | 0 | 50.76 | 49.32 |
| BLRM-SR | 60.11 | 27.14 | 12.43 | 12.52 | 12.53 | 0 | 30.27 | 29.41 |
| Scenario 3 | | | | | | | | |
| BARD-BOIN | 53.42 | 22.40 | 3.80 (3.51) | 3.77 (3.48) | 12.58 (12.61) | 0.76 (0.82) | 51.51 | 47.71 |
| BARD-BLRM | 50.07 | 22.00 | 6.26 (3.47) | 6.25 (3.51) | 12.57 (12.47) | 2.01 (0.82) | 32.74 | 30.68 |
| BOIN-SR | 65.56 | 29.78 | 12.61 | 12.52 | 12.58 | 0 | 50.29 | 47.86 |
| BLRM-SR | 64.96 | 29.84 | 12.52 | 12.50 | 12.43 | 0 | 29.26 | 27.63 |
| Scenario 4 | | | | | | | | |
| BARD-BOIN | 53.49 | 22.86 | 3.28 (3.52) | 3.32 (3.49) | 12.59 (12.56) | 0.65 (0.82) | 50.16 | 47.01 |
| BARD-BLRM | 50.92 | 22.69 | 5.05 (3.52) | 5.05 (3.5) | 12.74 (12.59) | 1.44 (0.83) | 30.97 | 29.47 |
| BOIN-SR | 64.75 | 29.43 | 12.57 | 12.51 | 12.57 | 0 | 48.02 | 45.56 |
| BLRM-SR | 65.21 | 30.03 | 12.63 | 12.41 | 12.59 | 0 | 29.59 | 28.22 |
| Scenario 5 | | | | | | | | |
| BARD-BOIN | 39.43 | 17.34 | 4.72 (3.50) | 4.80 (3.48) | 12.51 (12.6) | 1.06 (0.83) | 69.74 | 71.76 |
| BARD-BLRM | 32.36 | 14.92 | 7.68 (3.48) | 7.63 (3.51) | 12.67 (12.53) | 2.44 (0.83) | 52.86 | 54.97 |
| BOIN-SR | 54.60 | 24.56 | 12.61 | 12.55 | 12.55 | 0 | 66.68 | 66.96 |
| BLRM-SR | 46.45 | 21.39 | 12.69 | 12.66 | 12.62 | 0 | 51.74 | 52.05 |

| | N | | | | | | PCS1 | PCS2 |
|---|---|---|---|---|---|---|---|---|
| Scenario 6 | | | | | | | | |
| BARD-BOIN | 48.95 | 20.60 | 4.28 (3.52) | 4.25 (3.46) | 12.65 (12.60) | 0.91 (0.83) | 62.92 | 63.60 |
| BARD-BLRM | 44.35 | 19.43 | 6.50 (3.47) | 6.49 (3.49) | 12.70 (12.51) | 1.92 (0.82) | 67.51 | 67.18 |
| BOIN-SR | 63.30 | 28.42 | 12.58 | 12.53 | 12.54 | 0 | 59.40 | 60.37 |
| BLRM-SR | 60.11 | 27.14 | 12.52 | 12.57 | 12.58 | 0 | 64.35 | 65.88 |
| Scenario 7 | | | | | | | | |
| BARD-BOIN | 54.32 | 22.17 | 3.91 (3.52) | 3.89 (3.47) | 12.65 (12.64) | 0.79 (0.82) | 57.78 | 61.57 |
| BARD-BLRM | 50.88 | 21.72 | 6.39 (3.50) | 6.38 (3.49) | 12.63 (12.6) | 2.05 (0.82) | 60.30 | 63.31 |
| BOIN-SR | 65.56 | 29.78 | 12.61 | 12.52 | 12.58 | 0 | 54.53 | 57.50 |
| BLRM-SR | 64.96 | 29.84 | 12.52 | 12.61 | 12.60 | 0 | 56.75 | 60.83 |
| Scenario 8 | | | | | | | | |
| BARD-BOIN | 55.15 | 22.64 | 3.37 (3.54) | 3.40 (3.48) | 12.59 (12.52) | 0.67 (0.81) | 62.80 | 65.35 |
| BARD-BLRM | 52.34 | 22.35 | 5.11 (3.50) | 5.11 (3.47) | 12.75 (12.54) | 1.47 (0.82) | 52.66 | 54.70 |
| BOIN-SR | 64.75 | 29.43 | 12.57 | 12.51 | 12.57 | 0 | 62.42 | 65.38 |
| BLRM-SR | 65.21 | 30.03 | 12.44 | 12.64 | 12.62 | 0 | 50.11 | 53.16 |

N: average total sample size; PCS1: percentage of correct selection of the OBD based on the on the efficacy-rate-based approach; PCS2: the percentage of correct selection (PCS) of the true OBD based on the utility approach; *: numbers in paratheses are the results from the Pocock-Simon method with 40 patients randomized.

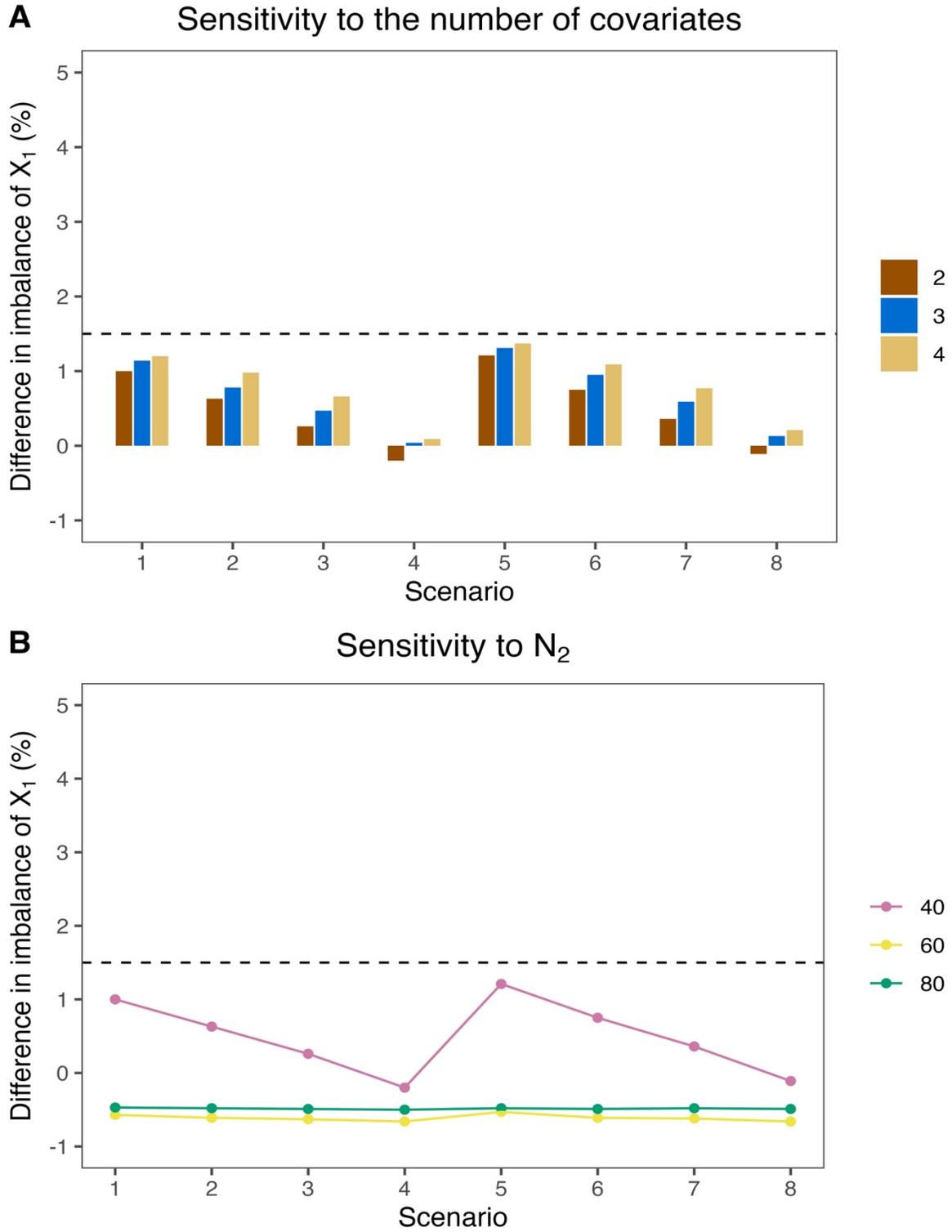

Figure 1. The difference in the imbalance index of $X_1$ between BARD-BOIN and Pocock-Simon method under different (A) numbers of covariates adjusted, and (B) sample size $N_2$ in stage 2.

# Supplementary Materials for "BARD: A seamless two-stage dose optimization design integrating backfill and adaptive randomization"

## 1. Details of simulation configurations

The response/efficacy rates of the simulation scenarios are generated by the following logistic regression model:

$$logit(p_E|d_j) = \beta_{0j} + \beta_1 * I(X_1 = 2) + \beta_2 * I(X_2 = 2) + \beta_3 * I(X_3 = 2),$$

where $\beta_1 = 1.7, \beta_2 = -1.5$ and $\beta_3 = 0.4$ across all the scenarios. The intercept $\beta_{0j}$ in each scenario is presented in Table S1.

Table S1. $\beta_{0j}$ in generating response rate.

| Scenario | Dose level | | | | |
|---|---|---|---|---|---|
| | 1 | 2 | 3 | 4 | 5 |
| 1 | -2.197 | -1.099 | -0.619 | -0.201 | 0.201 |
| 2 | -2.442 | -2.197 | -1.099 | -0.619 | -0.201 |
| 3 | -2.944 | -2.442 | -2.197 | -1.099 | -0.619 |
| 4 | -3.892 | -2.944 | -2.442 | -2.197 | -1.099 |
| 5 | -1.099 | -1.099 | -1.046 | -1.046 | -1.046 |
| 6 | -2.197 | -1.099 | -1.099 | -1.046 | -1.046 |
| 7 | -2.442 | -2.197 | -1.099 | -1.099 | -1.046 |
| 8 | -2.944 | -2.442 | -2.197 | -1.099 | -1.099 |

## 2. The rigidity of BLRM and the reasons for BARD-BLRM's notably worse covariate balance compared to BARD-BOIN

As described in the paper, BARD-BLRM shows notably worse covariate balance than BARD-BOIN, although both designs use the same covariate-adaptive randomization method in stage 2. A key factor contributing to this result is the rigidity of BLRM, stemming from the use of the two-parameter logistic model, which often causes the dose-finding process to become stuck at a particular dose. As a result, BLRM often leads to a highly imbalanced number of patients between $d_{low}$ and $d_{high}$ at the end of stage 1. This imbalance is carried over to stage 2, making it challenging to fully correct it given the limited sample size in that stage.

To illustrate this issue, Table S2 summarizes the following metrics related to stage 1:
- Average sample size of stage 1 ($N_1$).
- Percentage of correct selection (PCS) of stage 2 doses, which is the PCS of carrying forward the true OBD to stage 2 randomization.
- Average number of patients treated at $d_{low}$ ($n_{1,low}$) at the end of stage 1.
- Average number of patients treated at $d_{high}$ ($n_{1,high}$) at the end of stage 1.
- Imbalance of patient allocation between $d_{low}$ and $d_{high}$ at the end of stage 1.

As shown in Table S2, compared to BARD-BOIN, despite similar $N_1$, BARD-BLRM has a noticeably larger imbalance in patient allocation, making it challenging to fully correct given the limited sample size of stage 2. As a result, imbalance of BARD-BLRM is larger than BARD-BOIN in at the end of stage 2.

To further demonstrate the rigidity of BLRM, due to the overfitting issue of the two-parameter logistic model under small sample sizes, consider scenario 2 (in Table 3) where the true DLT rates for the five does are (0.04, 0.12, 0.25, 0.43, 0.63). Suppose that during the dose escalation process, the observed DLT data at the five doses are (0/3, 0/6, 2/3, -, -), where the denominator is the number of patients and the numerator is the number of patients with DLTs. Under the BLRM, the posterior probability of overdose for $d_3$ is 0.626, leading the dose to stay at $d_2$. If no DLT is observed at $d_2$ in subsequent patients, the dose-finding process becomes stuck at $d_2$, never exploring $d_3$. For example, even if the observe data are (0/3, 0/24, 2/3, -, -), indicating that $d_2$ is very safe, the dose-finding process remains stuck at $d_2$. This phenomenon occurs because the two-parameter logistic model overfits the (sparse) data at $d_3$. As a result, additional data at $d_2$ doesn't influence the DLT estimate at $d_3$.

Table S2. The imbalance of patient allocation at the end of stage 1 for BARD-BOIN and BARD-BLRM.

| Design | $N_1$ | PCS of stage 2 doses | $n_{1,low}$ | $n_{1,high}$ | Imbalance of allocation at stage 1 |
|---|---|---|---|---|---|
| Scenario 1 | | | | | |
| BARD-BOIN | 25.24 | 67.99 | 10.65 | 12.68 | 4.55 |
| BARD-BLRM | 23.43 | 59.44 | 8.83 | 17.32 | 9.76 |
| Scenario2 | | | | | |
| BARD-BOIN | 31.63 | 66.39 | 10.58 | 12.25 | 4.34 |
| BARD-BLRM | 29.31 | 40.65 | 8.15 | 16.03 | 8.83 |
| Scenario 3 | | | | | |
| BARD-BOIN | 35.94 | 70.43 | 10.96 | 11.52 | 3.91 |
| BARD-BLRM | 35.18 | 42.92 | 9.22 | 15.51 | 8.02 |
| Scenario 4 | | | | | |
| BARD-BOIN | 34.15 | 63.01 | 10.51 | 10.11 | 3.39 |
| BARD-BLRM | 34.21 | 38.40 | 9.00 | 13.88 | 6.74 |
| Scenario 5 | | | | | |
| BARD-BOIN | 26.40 | 88.98 | 12.03 | 12.70 | 3.93 |
| BARD-BLRM | 24.81 | 72.79 | 10.65 | 17.44 | 8.22 |
| Scenario 6 | | | | | |
| BARD-BOIN | 32.76 | 87.86 | 11.28 | 12.37 | 4.09 |
| BARD-BLRM | 30.42 | 90.28 | 8.87 | 16.30 | 8.43 |
| Scenario 7 | | | | | |
| BARD-BOIN | 37.54 | 81.56 | 11.59 | 11.60 | 3.72 |
| BARD-BLRM | 36.86 | 81.44 | 9.96 | 15.69 | 7.49 |
| Scenario 8 | | | | | |
| BARD-BOIN | 36.66 | 88.89 | 11.31 | 10.16 | 3.21 |
| BARD-BLRM | 36.66 | 71.99 | 9.94 | 14.02 | 6.03 |

3.  **Sensitivity analysis with 3 dose levels**

We evaluated the operating characteristics of BARD-BOIN with three dose levels and compared them to BOIN-SR in four scenarios, as shown in Table S4. The response/efficacy rates of the scenarios are generated by the following logistic regression model:

$$logit(p_E|d_j) = \beta_{0j} + \beta_1 * I(X_1 = 2) + \beta_2 * I(X_2 = 2) + \beta_3 * I(X_3 = 2),$$

where $\beta_1 = 1.7, \beta_2 = -1.5$ and $\beta_3 = 0.4$ across all the scenarios. The intercept $\beta_{0j}$ in each scenario is presented in Table S5.

The maximum number of stage 1 patients was set at 18. Both BARD-BOIN and BOIN-SR stop early if the number of patients at the current dose reaches $n_{stop} = 9$ and the decision is "stay". The simulation results are summarized in Table S6, and they are generally consistent with those observed with five doses. Specifically, compared to BOIN-SR, BARD-BOIN reduces the sample size and trial duration, achieves better covariate balance, and improves the accuracy of identifying the OBD.

Table S4. Simulation scenarios with 3 dose levels, with OBD highlighted in bold.

| Scenario | | Dose level | | |
|---|---|---|---|---|
| | | 1 | 2 | 3 |
| 1 | DLT | 0.12 | **0.25** | 0.40 |
| | Efficacy | 0.181 | **0.349** | 0.439 |
| | Utility | 38.6 | **45.2** | 44.4 |
| 2 | DLT | 0.04 | 0.12 | **0.25** |
| | Efficacy | 0.152 | 0.181 | **0.349** |
| | Utility | 39.3 | 38.6 | **45.2** |
| 3 | DLT | **0.12** | 0.25 | 0.42 |
| | Efficacy | **0.349** | 0.349 | 0.359 |
| | Utility | **50.0** | 45.2 | 39.5 |
| 4 | DLT | 0.04 | **0.12** | 0.25 |
| | Efficacy | 0.181 | **0.349** | 0.349 |
| | Utility | 41.3 | **50.0** | 45.2 |

Table S5. $\beta_{0j}$ used to generate response rate in scenarios of 3 dose levels.

| Scenario | Dose level | | |
|---|---|---|---|
| | 1 | 2 | 3 |
| 1 | -2.197 | -1.099 | -0.619 |
| 2 | -2.442 | -2.197 | -1.099 |
| 3 | -1.099 | -1.099 | -1.046 |
| 4 | -2.197 | -1.099 | -1.099 |

Table S6. Operating characteristics of BARD-BOIN for three dose levels, comparing to BOIN-SR.

| Design | N | Duration (month) | Imbalance* $X_1$ | Imbalance* $X_2$ | Imbalance* $X_3$ | Imbalance* allocation | PCS1 | PCS2 |
|---|---|---|---|---|---|---|---|---|
| Scenario 1 | | | | | | | | |
| BARD-BOIN | 37.79 | 16.44 | 2.96 (3.52) | 2.96 (3.46) | 12.58 (12.54) | 0.57 (0.82) | 50.67 | 48.10 |
| BOIN-SR | 50.08 | 21.85 | 12.64 | 12.54 | 12.5 | 0 | 51.46 | 50.00 |
| Scenario 2 | | | | | | | | |
| BARD-BOIN | 43.96 | 18.60 | 2.84 (3.48) | 2.80 (3.46) | 12.55 (12.43) | 0.53 (0.82) | 50.51 | 48.46 |
| BOIN-SR | 56.15 | 24.21 | 12.44 | 12.55 | 12.69 | 0 | 48.15 | 46.52 |
| Scenario 3 | | | | | | | | |
| BARD-BOIN | 37.81 | 16.06 | 3.03 (3.52) | 3.06 (3.46) | 12.59 (12.69) | 0.59 (0.83) | 68.16 | 69.96 |
| BOIN-SR | 50.08 | 21.85 | 12.64 | 12.54 | 12.5 | 0 | 65.24 | 65.60 |
| Scenario 4 | | | | | | | | |
| BARD-BOIN | 44.12 | 18.33 | 2.90 (3.46) | 2.85 (3.46) | 12.51 (12.47) | 0.55 (0.83) | 69.96 | 69.92 |
| BOIN-SR | 56.15 | 24.21 | 12.44 | 12.55 | 12.69 | 0 | 68.96 | 70.27 |

*: Numbers in paratheses are the results from the "full" Pocock-Simon method with 40 patients randomized.

### 4. Evaluation of estimation bias

Table S7 presents the biases of the estimates of the DLT rate at low dose ($\hat{p}_{low}$) and high dose ($\hat{p}_{high}$), where the estimates are the sample mean of the DLT rate at $d_{low}$ and $d_{high}$, respectively. As the primary objective of dose optimization involves the comparison between $d_{low}$ and $d_{high}$, the bias the estimate of the DLT rate difference between $d_{low}$ and $d_{high}$ (denoted as $\hat{\delta}_T = \hat{p}_{low} - \hat{p}_{high}$) is also reported. Similarly, the biases of the estimate of the response rate for low dose ($\hat{p}_{E,low}$), high dose ($\hat{p}_{E,high}$), and their difference (denoted as $\hat{\delta}_E = \hat{p}_{E,low} - \hat{p}_{E,high}$) are reported.

The results demonstrate that the estimate of the response rate has minimal bias. The estimate of the DLT rate has small negative bias (-0.015 to -0.021), but is generally negligible compared to the high heterogeneity of typical early phase patients and the variance of the DLT estimate. The estimates of BOIN-SR are unbiased because BOIN-SR uses only stage 2 randomization data.

Table S7. Bias of the estimate of the DLT rate for low dose ($\hat{p}_{low}$), high dose ($\hat{p}_{high}$), and their difference ($\hat{\delta}_T$); and the bias of the estimate of the response rate for low dose ($\hat{p}_{E,low}$), high dose ($\hat{p}_{E,high}$), and their difference ($\hat{\delta}_E$).

| Design | $\hat{p}_{low}$ | $\hat{p}_{high}$ | $\hat{\delta}_T$ | $\hat{p}_{E,low}$ | $\hat{p}_{E,high}$ | $\hat{\delta}_E$ |
|---|---|---|---|---|---|---|
| Scenario 1 | | | | | | |
| BARD-BOIN | -0.021 (0) | -0.004 (0) | -0.017 (0) | 0 (0) | 0 (0.003) | -0.001 (-0.003) |
| BOIN-SR | 0 | 0 | 0 | -0.001 | 0.002 | -0.003 |
| Scenario 2 | | | | | | |
| BARD-BOIN | -0.021 (0) | -0.001 (0.001) | -0.019 (-0.002) | -0.001 (0) | 0 (0) | 0 (0) |
| BOIN-SR | 0 | 0 | 0.001 | 0 | 0.001 | -0.001 |
| Scenario 3 | | | | | | |
| BARD-BOIN | -0.021 (-0.001) | 0 (0) | -0.021 (-0.001) | -0.001 (0) | 0 (0) | -0.001 (0) |
| BOIN-SR | 0 | 0 | 0.001 | 0 | 0 | 0 |
| Scenario 4 | | | | | | |
| BARD-BOIN | -0.015 (0) | 0.005 (0) | -0.020 (0) | -0.001 (-0.001) | 0 (0) | -0.001 (-0.001) |
| BOIN-SR | 0 | 0 | -0.001 | 0 | 0 | 0.001 |
| Scenario 5 | | | | | | |
| BARD-BOIN | -0.021 (0) | -0.004 (0.001) | -0.018 (-0.001) | 0 (0) | 0 (0) | 0 (0) |
| BOIN-SR | 0 | 0 | 0 | 0 | 0.001 | -0.001 |
| Scenario 6 | | | | | | |
| BARD-BOIN | -0.021 (0) | -0.002 (0.002) | -0.019 (-0.002) | 0 (0) | 0 (0) | 0 (0) |
| BOIN-SR | 0 | 0 | 0.001 | 0.001 | 0.001 | 0 |
| Scenario 7 | | | | | | |
| BARD-BOIN | -0.020 (-0.001) | 0.001 (0.001) | -0.021 (-0.001) | -0.001 (0.002) | 0 (0) | -0.001 (0.002) |
| BOIN-SR | 0 | 0 | 0.001 | 0 | 0.001 | 0 |
| Scenario 8 | | | | | | |
| BARD-BOIN | -0.015 (0) | 0.005 (0) | -0.020 (0) | -0.002 (0) | -0.001 (0.001) | -0.001 (-0.001) |
| BOIN-SR | 0 | 0 | -0.001 | 0.001 | 0 | 0.001 |

*: numbers in paratheses are the results from the Pocock-Simon method with 40 patients randomized.